# Coarse-Grained Molecular Dynamics Simulations of Oxidative Aging and Stabilization in Polymer Melts with Primary Antioxidants: Effects of Antioxidant Concentration and Molecular Architecture

*Takato Ishida[a]\*, Emmanuel Richaud[b]*

[a] Department of Materials Physics, Nagoya University, Furo-cho, Chikusa, Nagoya 464-8603, Japan

[b] Laboratoire PIMM, Arts et Metiers Institute of Technology, CNRS Cnam, HESAM Universite, 151 boulevard de l'Hopital, Paris 75013, France

\* E-mail: ishida@mp.pse.nagoya-u.ac.jp

**ABSTRACT**

   Industrial polymeric materials often rely on antioxidants to achieve long-term reliability. Previous studies have frequently discussed the stabilization effect in the presence of macroscopic additive migration. However, the micro- to meso-scale coupling between polymer dynamics and antioxidant molecular dynamics remains insufficiently understood. In this study, we extend a polymer dynamics simulation framework that can account for oxidative aging. We also update the model so that it can explicitly incorporate antioxidant molecules into the simulation. As a result, the framework enables us to quantify how molecular architecture of antioxidants affects oxidation kinetics, which has previously been inferred only indirectly from apparent changes in reaction rates. It also allows us to evaluate the effects of antioxidant concentration and molecular architecture on the spatial heterogeneity of oxidative aging.

**HIGHLIGHTS**

- Polymer oxidative-aging simulations incorporating primary antioxidants

- Effects of antioxidant concentration and molecular architecture on oxidation kinetics

- Mechanistic origin of architecture-dependent differences in stabilization efficiency

- Antioxidant concentration modulates the spatial heterogeneity of oxidative aging

## 1. INTRODUCTION

Stabilization with antioxidants is essential for extending the lifetime and ensuring the long-term reliability of polymeric materials. Antioxidants are used not only in industrial polymers but also in applications that require long-term stability, including medical devices such as artificial joints and syringes, packaging for foods and consumer products, and coatings used to protect artworks and cultural heritage materials [1–4]. The mechanisms by which antioxidants operate in polymer degradation based on the autoxidation reaction scheme are well understood [5]. There are two main classes of antioxidants: primary and secondary. In the autoxidation reaction, primary antioxidants—so-called chain-breaking antioxidants—scavenge peroxy radicals (POO·), whereas secondary antioxidants act as hydroperoxides (POOH) decomposers [6]. In addition, hindered amine light stabilizers (HALS) are widely used as light stabilizers; through a regenerative nitroxyl-radical cycle, they trap polymer radicals (P·) and thereby suppress photooxidative degradation. The retardation effect of antioxidants on oxidative aging has been quantitatively evaluated and validated through numerous chemical-kinetic modeling studies. Kinetic models have been proposed for a wide range of polymer systems, encompassing primary and secondary antioxidants as well as their synergistic interactions [7–12].

Regarding the concentration effect of antioxidants, it has been confirmed in various systems that the oxidation induction period is proportional to the concentration of phenolic groups, irrespective of the antioxidant type or the polymer being stabilized [7,8,13]. Breeze et al. [14] investigated the stabilization performance of a range of hindered phenolic antioxidants with different chemical structures, using squalane as a model product whose oxidation mechanism is considered to be similar to that of polypropylene (PP). Within the 2,6-di-tert-butylphenol family, higher-molecular-weight hindered phenols (e.g., Irganox1010) were reported to provide a more significant stabilization effect than dibutylhydroxytoluene (BHT) in squalane–phenolic antioxidant systems under thermal oxidation at 190 °C. However, in experimental evaluations based on oxidation kinetics, the effect of evaporation of low-molecular-weight antioxidants [11,15,16], such as BHT, cannot be eliminated. To gain a deeper understanding of the intrinsic nature of antioxidant additives without the influence of evaporation, a simulation approach that explicitly treats and tracks the diffusion of antioxidant molecules is effective.

In this study, we extend a coarse-grained molecular dynamics (CGMD) framework [17–19] that explicitly simulates the dynamics of polymers and macromolecular radicals during oxidative aging to incorporate primary antioxidants. We investigate the molecular-architecture effect of antioxidants using polymer-melt systems containing two types of coarse-grained (CG) antioxidant molecules, designed to mimic BHT and Irganox1010. We also systematically vary the antioxidant concentration. We perform oxidative aging simulations under these conditions and discuss how the concentration affects the oxidative aging kinetics and the spatial heterogeneity of aging.

## 2. MODEL AND SIMULATIONS

CGMD simulations are used to examine the oxidative aging behavior of polymers, following methods established in previous papers [17–19]. To incorporate the stabilizing effect of primary antioxidants through scavenging POO· radicals, we update the previous framework [17,19]. For polymer dynamics, we employ the Kremer–Grest (KG) model [20], which accounts for excluded-volume interactions between polymer segments and the finite extensibility of bonds. The total potential energy is given by the sum of the Weeks–Chandler–Andersen (WCA) potential, $U_{\text{WCA}}$, which represents excluded-volume interactions acting between all particles, and the finite extensible nonlinear elastic (FENE) spring potential, $U_{\text{FENE}}$, which represents the finite extensibility of bonds. The potentials $U_{\text{WCA}}$ and $U_{\text{FENE}}$ are expressed as follows.

$$U_{\text{FENE}}(r) = \begin{cases} -0.5 k R_0^2 \ln\left[1 - \left(\frac{r}{R_0}\right)\right], & r \leq R_0 \\ \infty, & r > R_0 \end{cases} \quad (1)$$

$$U_{\text{WCA}}(r) = \begin{cases} 4\varepsilon \left[\left(\frac{\sigma}{r}\right)^{12} - \left(\frac{\sigma}{r}\right)^{6} + \frac{1}{4}\right], & r \leq 2^{1/6}\sigma \\ 0, & r > 2^{1/6}\sigma \end{cases} \quad (2)$$

Here, $r$ is the distance between two beads. The FENE parameters are $k$ (spring constant) and $R_0$ (maximum bond length). The WCA parameters are $\varepsilon$ and $\sigma$, where $\varepsilon$ sets the energy scale and $\sigma$ sets the length scale (bead diameter). Unless otherwise noted, we adopt the standard KG settings [20]: $R_0 = 1.5\sigma, k = 30\varepsilon/\sigma^2, \rho^* = 0.85/\sigma^3, k_B T = 1.0\varepsilon$. In this context, $\rho^*$ denotes the number density of segment beads, and this value corresponds to a typical density of polymer melts. $k_B$ is the Boltzmann constant and $T$ is the temperature. The equation of motion of beads follows the Langevin equation, as shown below.

$$m \frac{d^2 \boldsymbol{r}_i}{dt^2} = -\frac{\partial U_{\text{tot}}}{\partial \boldsymbol{r}_i} - \Gamma \frac{d\boldsymbol{r}_i}{dt} + \boldsymbol{W}_i(t) \quad (3)$$

In the Langevin equation, $m$ denotes the bead mass and $\mathbf{r}_i$ is the position of bead $i$. We set the friction coefficient to $\Gamma = 1.0$. The thermal noise term $\mathbf{W}_i(t)$ is modeled as Gaussian white noise, obeying $\langle \mathbf{W}_i(t) \rangle = 0$ and $\langle \mathbf{W}_i(t)\mathbf{W}_j(t') \rangle = 2k_B T \Gamma \delta_{ij} \delta(t-t') \mathbf{I}$. In this notation, $\langle \cdots \rangle$ denotes a statistical average and $\mathbf{I}$ is the unit tensor. The total potential $U_{\text{tot}}$ is given by:

$$U_{\text{tot}} = \sum_{i>j} U_{\text{WCA}}(r_{ij}) + \sum_{i,j \in S_{\text{bond}}} U_{\text{FENE}}(r_{ij}) \qquad (4)$$

The bead–bead distance is given by $r_{ij} = |\mathbf{r}_i - \mathbf{r}_j|$, and the set of bonded pairs are specified by $S_{\text{bond}}$. We use reduced units with mass $m$, length $\sigma$, and energy $\varepsilon$, and define the time unit as $\tau = \sigma\sqrt{m/\varepsilon}$. In the present simulations we set $m = \sigma = \varepsilon = 1$ (thus $\tau = 1$), and use a time step of $\Delta t = 0.005$ in the NVT ensemble for all simulations. For the chemical reactions of oxidative aging, we follow the approach established in previous papers [17–19] and update it to incorporate the effects of primary antioxidants in this study. Specifically, the framework employs a kinetic model for PP oxidative aging [21,22] based on the autoxidation scheme proposed by Bolland and Gee [23]. Here, we consider the following two POO· radical scavenging reaction mechanisms for primary antioxidants, which were previously concluded to be reasonable based on our previous work [8].

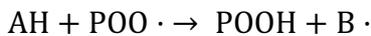

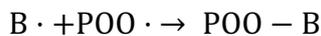

$$\text{AH} + \text{POO}\cdot \rightarrow \text{POOH} + \text{B}\cdot$$
$$\text{B}\cdot + \text{POO}\cdot \rightarrow \text{POO}-\text{B}$$

In this scheme, AH denotes a phenolic group, POOH is a hydroperoxide, and B· represents a cyclohexadienonyl radical. The cyclohexadienonyl radical is formed by isomerization of a phenoxy radical, and we assume that this isomerization process is sufficiently fast. Although phenolic antioxidants can in principle follow multiple competing reaction channels (e.g., side reactions of phenoxy-derived radicals and subsequent products), our goal here is to capture the dominant "chain-breaking" contribution in a minimal form; therefore, we lump these secondary processes into the effective termination step shown above, rather than introducing a more highly parameterized kinetic network. The terminated chemical structure of the resulting POO–B species is assumed to be a cyclohexadienone [7]. It is known that scavenging of POO· by the cyclohexadienonyl radical proceeds at a comparable rate to scavenging of POO· by the phenolic group [8].

In this study, to investigate the molecular-size effect on the stabilization performance of primary antioxidants, we model two types of CG antioxidant molecules. One is a single-bead molecule that represents BHT. The other is a five-bead molecule that represents Irganox1010 and consists of one unreactive linker bead connected to four antioxidant beads that provide the four reactive functionalities. Figure 1 summarizes the POO· scavenging reactions considered in this study and the corresponding reaction topologies. In this framework, chemical reactions are treated using the REACTION package [24,25] implemented in LAMMPS (Large-scale Atomic/Molecular Massively Parallel Simulation) [26], where reaction events are determined probabilistically. A bimolecular reaction may occur only when two beads are within the reaction cutoff distance, $r_c = 2^{1/6}$, and when the local configuration satisfies the required reaction topology, as shown in Fig. 1, as well as the autoxidation reaction topologies for PP reported previously [17–19].

**Radical scavenging reactions**

(i) AH + POO· → POOH + B·

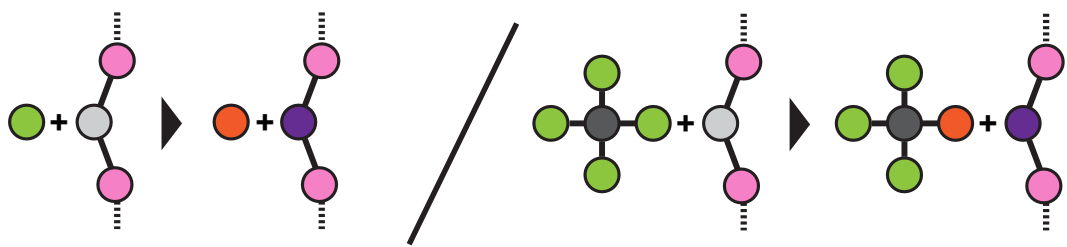

(ii) B· + POO· → POO-B

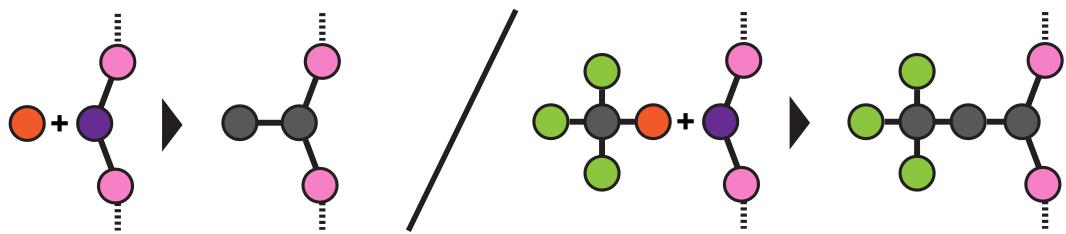

Figure 1 Schematic representation of the coarse-grained antioxidant models with two distinct molecular topologies (BHT-type and Irganox1010-type) and the corresponding reaction topology. AH (antioxidant group): green, B· (cyclohexadienonyl radical; isomerized phenoxy radical): orange, PH (polymer substrate): pink, POOH (hydroperoxide): purple, POO· (peroxy radical): white, unreactive beads: gray.

In this study, we investigate four initial antioxidant group concentrations, $[AH]_0$ = 0.2, 0.1, 0.02, and 0.002 mol/L, for both BHT-type and Irganox1010-type antioxidants. The polymer melt consists of 2560 KG chains with a degree of polymerization of N=100. We introduce a number of CG antioxidant molecules corresponding to each $[AH]_0$, and we determine the simulation box size so that the total number density satisfies $\rho^* = 0.85$. For each $[AH]_0$ = 0.2, 0.1, 0.02, and 0.002 mol/L, we introduced 2000, 100, 200, and 20 BHT-type antioxidant beads, respectively, and 500, 250, 50, and 5 Irganox1010-type antioxidant units, respectively, based on the spatiotemporal mapping of the KG model [27]. We here describe how the number of antioxidant beads was determined in the simulations. In the unstabilized PP aging system, we assume $[PH]_0$ = 20.3 mol/L, and $[POOH]_0 = 10^{-4}$ mol/L. With the present system size, this corresponds to initially containing one POOH species. In the stabilized systems, we set the initial number of antioxidant beads based on this reference $[POOH]_0$:

specifically, the number of introduced AH groups was chosen so that the simulation satisfies the prescribed ratio $[AH]_0/[POOH]_0$, i.e., the total AH beads count scales as $[AH]_0/[POOH]_0$ relative to the initial POOH count. In the reduced unit system, the considered values of $[AH]_0$ correspond to $6.6 \times 10^{-3}, 3.3 \times 10^{-3}, 6.6 \times 10^{-4}$, and $6.6 \times 10^{-5}$.

We primarily discuss how the antioxidant concentration affects the oxidative aging kinetics and the spatial heterogeneity of aging. Here, we investigate the effect of adding primary antioxidants to an unstabilized polymer melt system that is known to exhibit heterogeneous aging, where the average frequency of H-abstraction exceeds the chain relaxation rate (i.e., hydrogen abstraction occurs faster than chain relaxation) [17]. In this work, we set the H-abstraction rate constant to $k_3 = 5.0 \times 10^{-2}/\tau$ where $\tau$ is the simulation time unit. All other rate constants are set according to their ratios to $k_3$, specifically $k_{1u}/k_3 = 1.6 \times 10^{-4}$ and $k_{1b}/k_3 = 5.8 \times 10^{-3}$, based on a kinetic model for PP subjected to thermal oxidative aging at 180 °C [21]. For simplicity, we here assume that the conversion of P·) to POO· proceeds rapidly via oxygen addition in the autoxidation mechanism. The H-abstraction rate is fixed and determines the overall reaction rate. We then perform oxidative aging simulations by introducing BHT-type and Irganox1010-type antioxidants at the numbers corresponding to the aforementioned $[AH]_0$ = 0.2, 0.1, 0.02, and 0.002 mol/L. The radical-scavenging reaction rate of antioxidant groups has been reported in an experimental kinetic model [8], and applying those values to the present conditions indicates that scavenging is much faster than H-abstraction and POOH decomposition. Therefore, we set the scavenging reactions to occur immediately once the reaction topology in Fig. 1 is satisfied at a given time step. All simulations in this study were carried out using LAMMPS (version 23Jun22) [26]. We first fully equilibrated the polymer melt containing antioxidant molecules (BHT-type or Irganox1010-type). We then introduced a single free radical to initiate the oxidative aging simulations.

## 3. RESULTS AND DISCUSSION

We first discuss the chemical kinetics of primary antioxidants. Figure 2 shows the consumption of PH and AH, representing the overall oxidative aging behavior, for different antioxidant types and initial antioxidant-group concentrations, $[AH]_0$. Consistently with experimental observations for the oxidation in the olefinic family [21–23], our simulations also confirm that the autocatalytic oxidative aging behavior with an induction period is preserved even in the presence of antioxidants. In addition, Fig. 2 shows that the induction period ends only after the antioxidant groups (AH) begin to deplete, after which PH consumption accelerates. Qualitatively, this trend is consistent with the common understanding that the onset of the accelerated stage follows antioxidant depletion. However, in many experimental systems the induction period is often considered to end only after the antioxidant is nearly or fully consumed [28,29], whereas this is not strictly the case in the present simulations. This discrepancy may be related to the aging heterogeneity observed in our simulations, which we discuss in the latter part of this section. Note that the induction period and its concentration dependence appear to vary with the antioxidant type. The kinetic curves in Fig. 2 show scatter in the induction period due to the stochastic nature of the chemical reactions. Therefore, Fig. 3(left) presents the induction period ($t_{\text{ind}}$) and reaction time ($t_{\text{react}}$) averaged over 8 independent simulations initiated from different initial configurations and using different random seeds. The conversion $\alpha$ is defined from the depletion of PH beads as $\alpha = 1 - \frac{N_{\text{PH}}(t)}{N_{\text{PH}_0}}$, where $N_{\text{PH}}(t)$ is the number of PH beads at time $t$ and $N_{\text{PH}_0}$ is the initial value. For each simulation, $t_{\text{ind}}$ is determined from the low-conversion region of the PH-decay kinetic curve shown in Fig. 2. Following our previous work [19], $t_{\text{ind}}$ is obtained by linear extrapolation of the conversion–time relation using the two data points at $\alpha = 0.2$ and $\alpha = 0.3$; $t_{\text{ind}}$ is then defined as the intercept time with the baseline (at $\alpha = 0$). Using this definition, we further define the reaction time $t_{\text{react}}$ as the time interval from $t_{\text{ind}}$ to the time at which the conversion reaches $\alpha = 0.5$, i.e., $t_{\text{react}} = t(\alpha = 0.5) - t_{\text{ind}}$. The established proportionality between the initial antioxidant group concentration [7,8] and $t_{\text{ind}}$ is also confirmed in our simulations (Fig. 3(right)). Notably, the slope of the linear $[AH]_0$–$t_{\text{ind}}$ relation is larger for the BHT-type antioxidant, suggesting that BHT-type additives are more effective than Irganox1010-type antioxidants in extending $t_{\text{ind}}$. This behavior differs from the results reported in a previous study [8,14]. We attribute this difference mainly to the absence of antioxidant volatilization in our simulations.

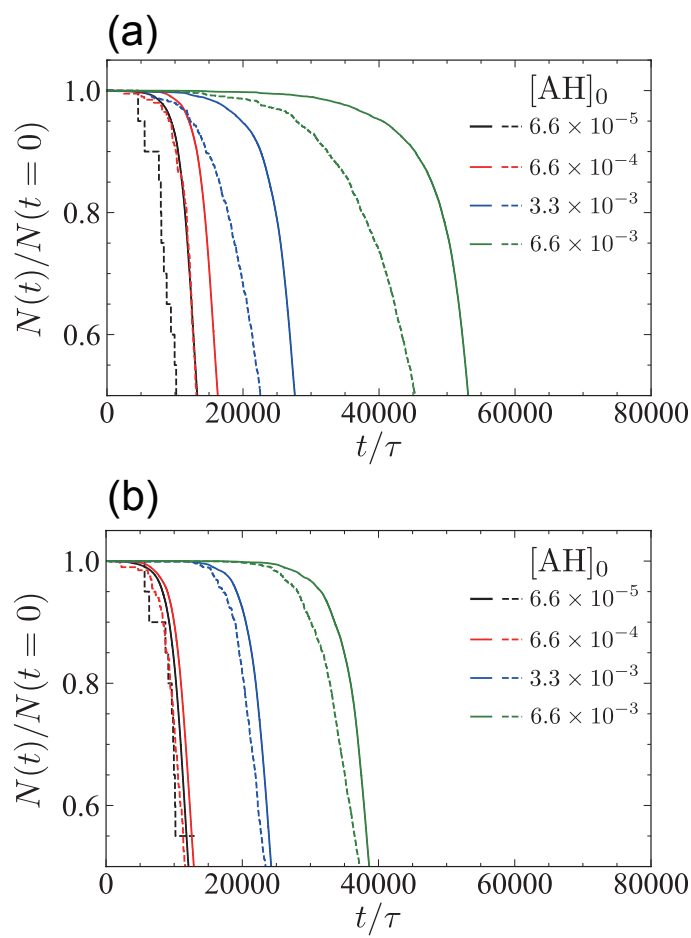

Figure 2. Kinetic curves of PH (polymer substrate) and AH (antioxidant group) decay as a function of aging time for $[AH]_0 = 6.6 \times 10^{-3}, 3.3 \times 10^{-3}, 6.6 \times 10^{-4}$ and $6.6 \times 10^{-5}$. (a) BHT-type antioxidant. (b) Irganox1010-type antioxidant. Solid lines denote PH, and dotted lines denote AH.

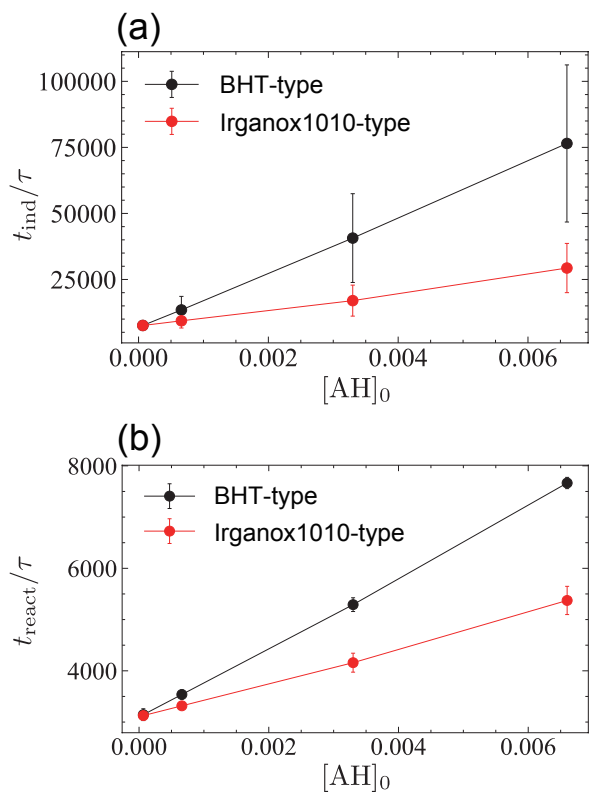

Figure 3 (a) Induction period $t_{\text{ind}}$ and (b) reaction time $t_{\text{react}}$ as a function of the initial antioxidant-group concentration $[AH]_0$ for the BHT-type and Irganox1010-type antioxidants.

The induction period ends when POOH accumulates and reaches a certain level of concentration, after which the oxidation reaction enters the accelerated stage. In our simulations, a few POO· radicals hop between molecules via H-abstraction reactions and leave a trail of POOH along their hopping paths on chains. The POO·-scavenging mechanism of antioxidants markedly reduces the number of radicals in the system, and in some cases eliminates them entirely, thereby delaying the accumulation of POOH, as shown in Fig. 4. In the unstabilized system, POOH is known to scale linearly with time (POOH $\propto t^1$) in the short-time regime before $t_{\text{ind}}$. In systems with a low $[AH]_0$ that show only limited stabilization, the linear relationship POOH $\propto t^1$ for $t < t_{\text{ind}}$ is maintained. In contrast, in systems with a high $[AH]_0$ that exhibit pronounced stabilization, this relationship breaks down and POOH displays a plateau over a certain time domain. To discuss the retardation effect during the accelerated stage ($t > t_{\text{ind}}$), Fig. 5 shows the evolution of the number of POO· radicals in the accelerated stage for various stabilized systems. Even when plotted against $t - t_{\text{ind}}$, the increase in POO· becomes slower as [AH]$_0$ increases. In addition, the BHT-type antioxidant mitigates the growth of POO· more effectively than the Irganox1010-type antioxidant. When [AH]$_0$ is low, the growth behavior of POO· is almost indistinguishable from that of the reference unstabilized system (i.e., [AH]$_0 = 0$) shown in Fig. 5. This suggests that the antioxidant concentration is below the minimum level required to manifest a measurable stabilization effect, and may be related to the critical antioxidant concentration (i.e., the threshold concentration) that has been discussed in the literature [30,31]. At [AH]$_0 = 6.6 \times 10^{-4}$, the POO· evolution in the Irganox1010-type system essentially overlaps with that of the unstabilized system, whereas the BHT-type system still shows a slight stabilization effect. This difference likely reflects not only the higher mobility of the BHT-type antioxidant but also differences in the local coordination environment arising from the antioxidant molecular architecture, as discussed below.

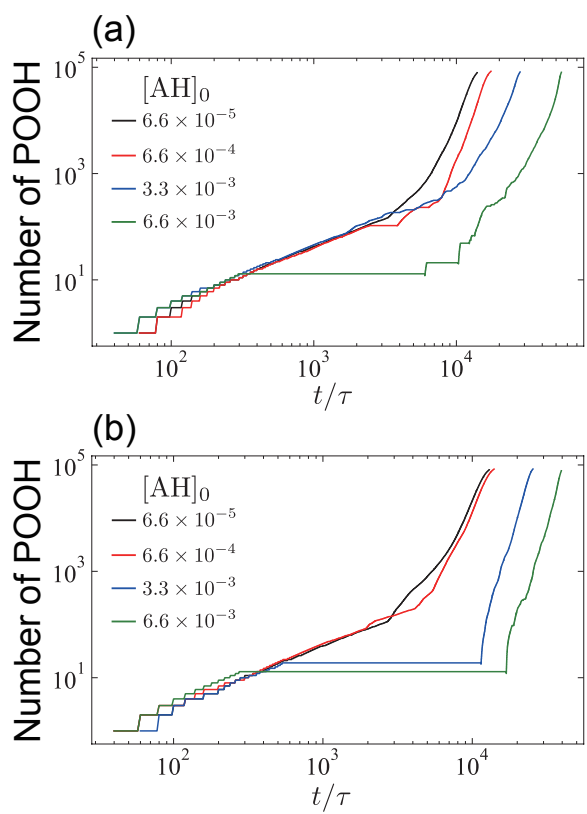

Figure 4 Evolution of POOH as a function of aging time for [AH]$_0 = 6.6 \times 10^{-3}, 3.3 \times 10^{-3}$, $6.6 \times 10^{-4}$ and $6.6 \times 10^{-5}$. (a): BHT-type antioxidant. (b): Irganox1010-type antioxidant.

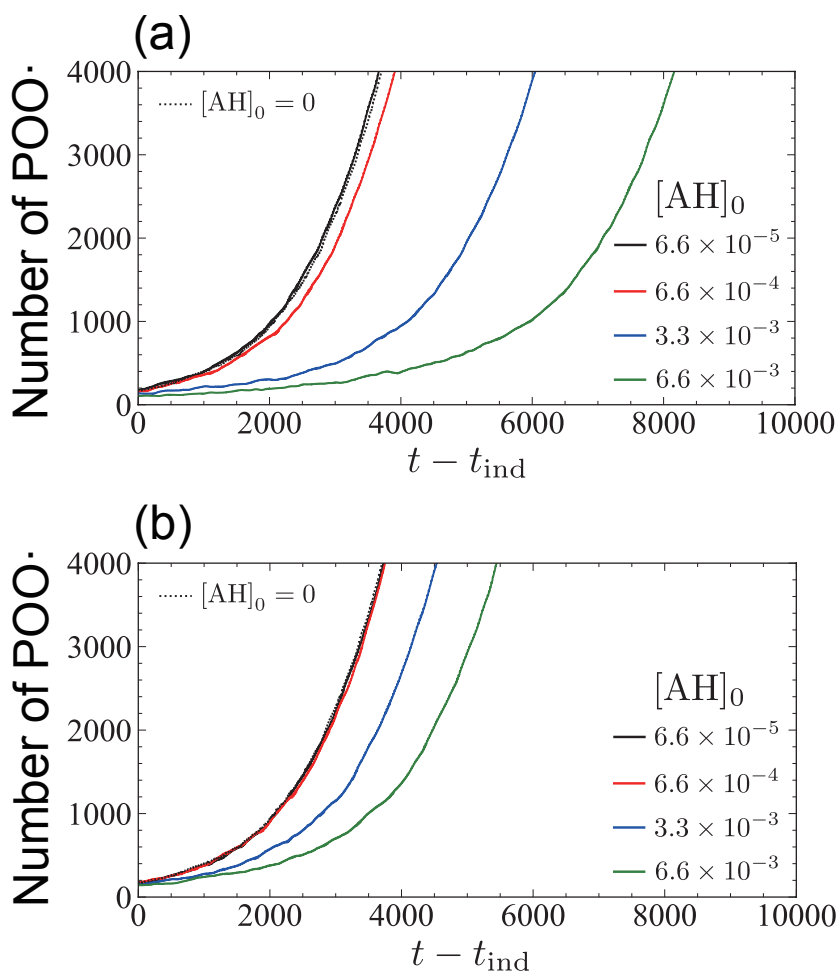

Figure 5 Evolution of POO· as a function of $t - t_{\text{ind}}$ for various stabilized systems. (a) BHT-type antioxidant. (b) Irganox1010-type antioxidant. For reference, the unstabilized system without antioxidants ($[AH]_0 = 0$) is also shown as a black dotted line.

The difference in POO· scavenging efficiency between the BHT-type and Irganox1010-type antioxidants is likely governed not only by their different diffusivities, which mainly reflect the molecular-size difference, but also by contributions from the static local structure around the antioxidant molecules. To examine the molecular mobility of the antioxidants, we plot in Fig. 6 the mean-squared displacement (MSD) of the center of mass of each antioxidant molecule. Both antioxidants diffuse consistently faster than the polymer segments shown for reference. As expected, the smaller BHT-type antioxidant exhibits higher mobility than the bulkier Irganox1010-type antioxidant. In the long-time Fickian diffusion regime, the diffusion coefficient can be estimated from the linear relation $\text{MSD} = 6Dt$. From this analysis, we obtain $D_{\text{BHT}} = 3.36 \times 10^{-2}$ and $D_{\text{Irganox1010}} = 8.57 \times 10^{-3}$, indicating that the BHT-type antioxidant diffuses approximately four times faster than Irganox1010-type antioxidant. This difference in mobility is therefore expected to contribute to the observed difference in POO· scavenging performance. In the following, we discuss the static local structure surrounding the antioxidant molecules. Figure 6 shows the radial distribution function $g(r)$ between PH and AH. In this simulation framework, only reactive beads located within the reaction cutoff distance $r_c$ contribute to chemical reactions, i.e., those in the short-distance region $r < r_c$. Using the results in Fig. 7, the coordination number can be evaluated from $N(r_c) = 4\pi\rho \int_0^{r_c} dr g(r) r^2$. The resulting coordination numbers for the BHT-type and Irganox1010-type systems are 3.44 and 2.46, respectively. Thus, the intrinsic molecular topology plays a role: the Irganox1010-type antioxidant exhibits a smaller coordination number with polymer segments. As a result, even when a fraction of polymer segments becomes radicalized, the probability that these radicals encounter and are scavenged by an Irganox1010-type antioxidant group is lower than that for the BHT-type antioxidant.

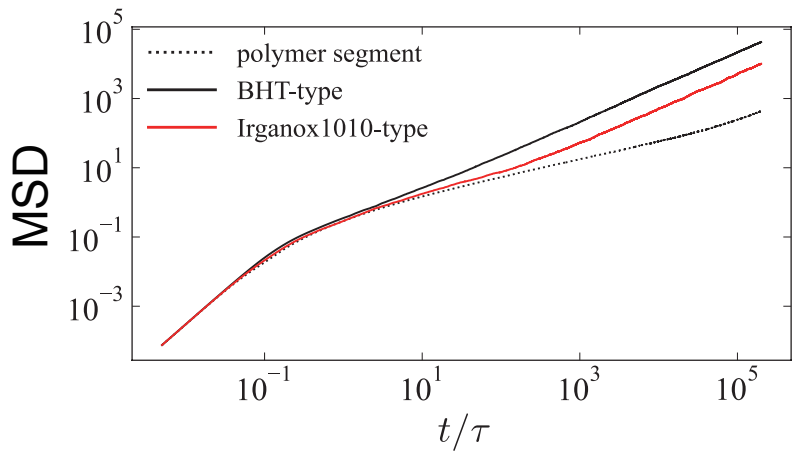

Figure 6 Mean-squared displacement (MSD) of the center of mass for BHT-type and Irganox1010-type antioxidants. The black dotted line shows the MSD of polymer segments and is included for reference.

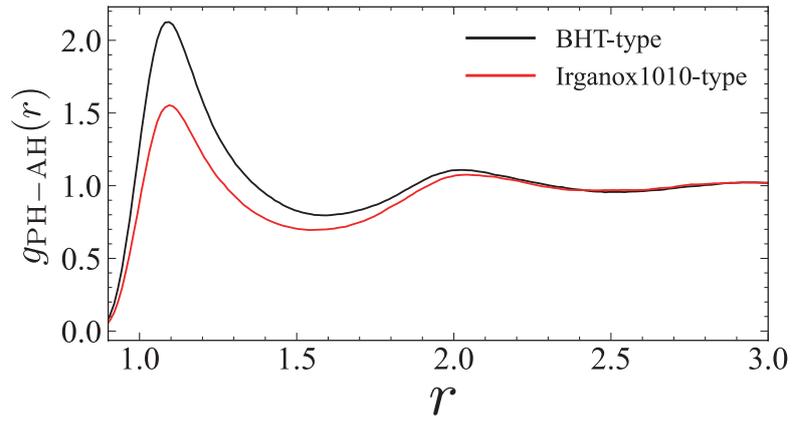

Figure 7 Radial distribution functions $g(r)$ between PH and AH for BHT-type and Irganox1010-type antioxidants at $[AH]_0 = 6.6 \times 10^{-3}$.

Figure 8 presents xy-plane projections of representative snapshots from oxidative aging simulations for two antioxidant types (BHT-type and Irganox1010-type) at two initial antioxidant group concentrations, $[AH]_0 = 6.6 \times 10^{-3}$ and $6.6 \times 10^{-5}$. At the low antioxidant group concentration $[AH]_0 = 6.6 \times 10^{-5}$, the spatially heterogeneous structure of the aged region shows little difference between the BHT-type and Irganox1010-type systems. In contrast, regardless of antioxidant type, increasing the antioxidant-group concentration to $[AH]_0 = 6.6 \times 10^{-3}$ mitigates the heterogeneity to some extent.

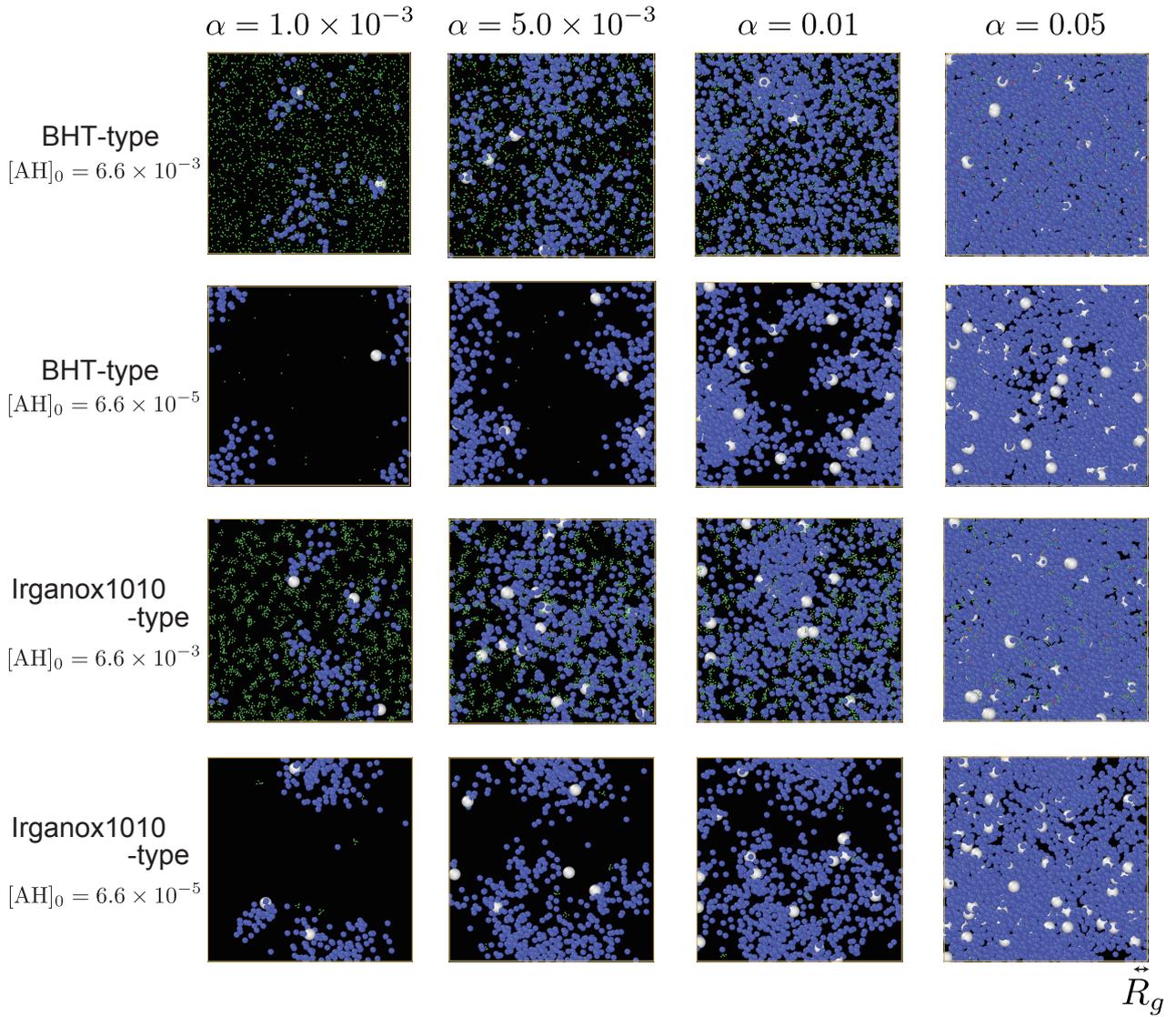

Figure 8 Representative snapshots from oxidative aging simulations at conversions $\alpha\left(= 1 - \frac{N_{\mathrm{PH}}(t)}{N_{\mathrm{PH}_0}}\right) = 1.0 \times 10^{-3}, 5.0 \times 10^{-3}, 0.01,$ and $0.05$ for the two antioxidant types (BHT-type and Irganox 1010-type) and two initial antioxidant group concentrations, $[AH]_0 = 6.6 \times 10^{-3}$ and $6.6 \times 10^{-5}$. Antioxidant groups (AH) are shown in green, cyclohexadienonyl radicals (B·) in red, unreactive beads in gray, peroxy radicals (POO·) in white, and scission sites in blue. Black regions indicate unreacted polymer. The scale bar corresponds to the initial radius of gyration $R_g$ of the polymer chains.

To discuss how the concentration and type of antioxidant mitigate the spatial heterogeneity of oxidative aging, we calculate the structure factor $S(q)$ from the spatial distribution of scission sites

(blue beads in Fig. 8). $S(q)$ is defined as:

$$S(q) = \left\langle \frac{1}{M} \sum_{i=1}^{M} \sum_{j=1}^{i-1} \exp(i\boldsymbol{q} \cdot \boldsymbol{r}_{ij}) \right\rangle \qquad (5)$$

where $M$ represents the number of the scission-end beads in the calculation. In this paper, $S(q)$ is obtained as the ensemble average of 8 independent simulation runs using snapshots at corresponding time points for each conversion $\alpha$. Figure 9(top) presents $S(q)$ at $\alpha = 0.1$ for the BHT-type and Irganox1010-type systems at different initial antioxidant group concentrations $[AH]_0$. The upturn of $S(q)$ in the low-$q$ region indicates the presence of long-wavelength concentration fluctuations of scission sites, i.e., spatial heterogeneity. A higher $[AH]_0$ suppresses the development of heterogeneous structures. The mitigation of heterogeneous structures observed at high $[AH]_0$ was more pronounced for the BHT-type antioxidant than for the Irganox1010-type antioxidant. At $\alpha = 0.4$, the $S(q)$ profiles become nearly insensitive to antioxidant type and concentration (Fig. 9, bottom). This convergence is consistent with a regime where the stabilization effect has largely diminished due to advanced antioxidant consumption and chain scission–accelerated relaxation makes the heterogeneity fuzzier.

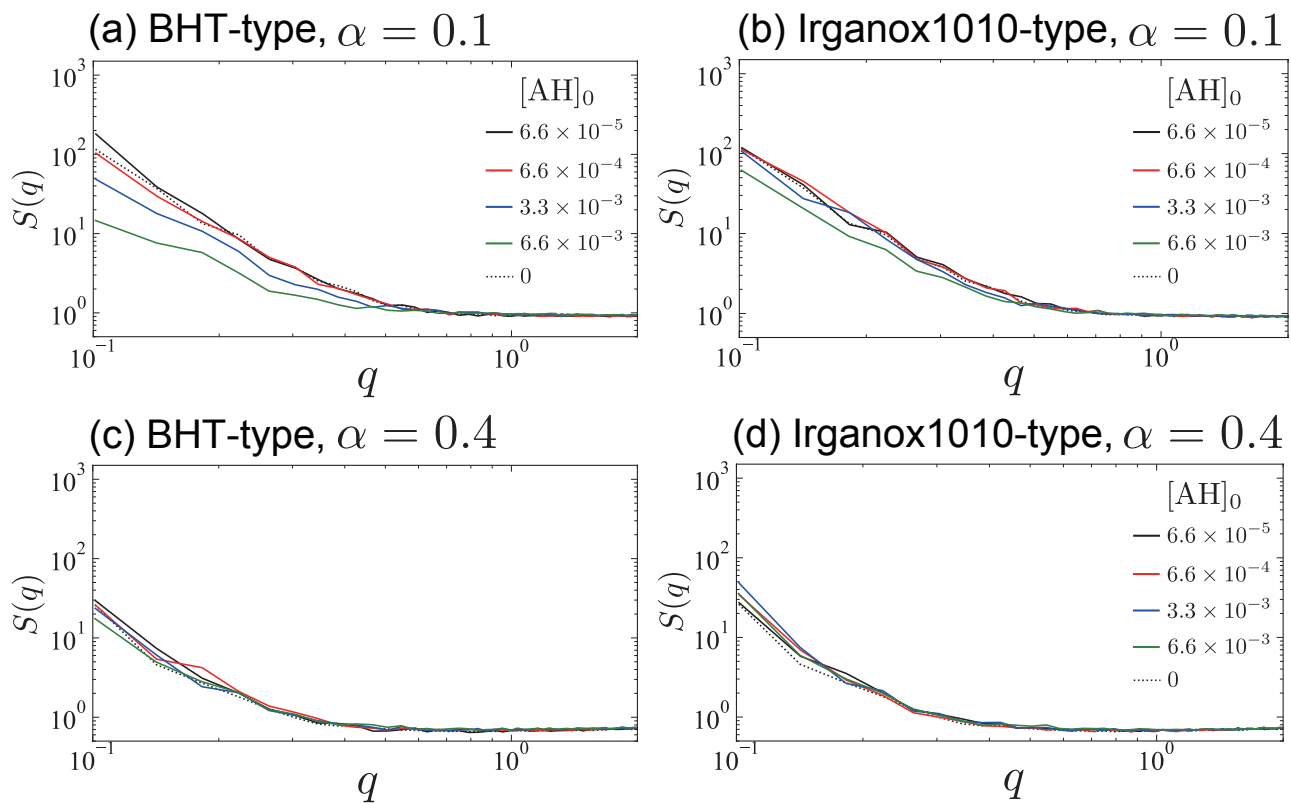

Figure 9. Structure factor $S(q)$ of scission-end beads in polymer melts containing two primary antioxidant types at different initial antioxidant-group concentrations, $[AH]_0$. (a, c) BHT-type. (b, d): Irganox1010-type. Results are shown at low conversion ($\alpha = 0.1$, top) and high conversion ($\alpha = 0.4$, bottom). For reference, the unstabilized system without antioxidants ($[AH]_0 = 0$) is also shown as a black dotted line.

For reference, the scission-end structure factors $S(q)$ of the unstabilized system are also shown in Fig. 9. At the low conversion $\alpha = 0.1$, we find that, for both antioxidant types, the systems with low $[AH]_0$ (ca. $6.6 \times 10^{-4}$ or lower) show essentially the same $S(q)$ as the unstabilized system. In contrast, at the higher conversion $\alpha = 0.4$, the influence of antioxidant type and concentration on the spatial heterogeneity becomes negligible, and the resulting $S(q)$ collapse onto that of the unstabilized system. Spatial heterogeneity during oxidative aging is governed by the competition between oxidation reactions—more specifically, the H-abstraction reactions—and chain relaxation, as shown in our previous study [17]. In the present study, antioxidants delay the reactions by scavenging radicals. Therefore, at higher [AH]$_0$, the oxidation proceeds more slowly, and the heterogeneous structure has more time to relax even if the intrinsic dynamics of polymer chains and radicals are unchanged. At high conversion, the decrease of $S(q)$ at low $q$ is consistent with progressive diffusion-driven homogenization of the heterogeneous structure, promoted by scission-accelerated relaxation. Consequently, it is not surprising that, once aging proceeds sufficiently, the remaining spatial heterogeneity becomes largely independent of whether antioxidants are present. At low conversion, we still observe clear effects of antioxidant type and concentration. However, when the antioxidant-group concentration is low, the heterogeneity becomes essentially the same as in the unstabilized system. This suggests the existence of a critical antioxidant concentration below which antioxidants no longer mitigate spatial heterogeneity. This trend mirrors the POO· evolution shown in Fig. 5. Thus, the commonly discussed concept of a critical antioxidant concentration may apply not only to chemical kinetics but also to the spatial heterogeneity of oxidative aging. Evaporation can also alter the macroscopic mode of heterogeneous oxidation at high antioxidant loadings, as discussed experimentally by Gijsman and Fiorio [32]. Because such loss pathways are absent in the present model, we focus here on the non-volatile contribution of antioxidants.

One may argue that the present simulations do not strictly follow the commonly assumed picture in the polymer aging/stabilization literature [28,29] where the induction period ends only after near-complete depletion of antioxidants. In our system, accelerated PH consumption can begin while a non-negligible fraction of AH still remains. A plausible reason is the enhanced bimolecular decomposition of POOH which arises from the spatially heterogeneous nature of oxidative aging in our model. In the present framework, reaction events are sampled using rate parameters mapped to the macroscopic oxidation kinetic model. However, in a spatially heterogeneous system, local encounter rates for bimolecular reactions do not necessarily remain consistent with the overall (well-stirred) kinetic description. In our simulations, scission sites, radicals, and POOH tend to localize within aged regions. This nonuniform spatial distribution biases the system toward more frequent POOH–POOH encounters and thereby promotes bimolecular POOH decomposition. As a consequence, localized "hot spots" with elevated radical concentrations emerge, where radical scavenging by antioxidants and autooxidation reactions proceed concurrently. This concurrency naturally leads to the behavior observed in Fig. 2, namely that AH consumption and the progression of autooxidation overlap in time, rather than showing a strictly sequential scenario in which antioxidant depletion clearly precedes the onset of accelerated oxidation. We also observe that the AH-consumption profiles differ between the BHT-type and Irganox1010-type antioxidants. This difference likely reflects distinct effective scavenging rates that depend on the dynamics and local coordination structure of antioxidants (see Figs. 6 and 7), which control how efficiently antioxidant groups encounter and scavenge radicals in the heterogeneous environment. These considerations further imply that, under homogeneous aging case (e.g., at reduced oxidation rate), the overlap between antioxidant consumption and autooxidation would be less pronounced. In that regime, one may expect to recover the more widely assumed scenario observed experimentally, in which the induction period ends only after antioxidants have been almost fully depleted.

Finally, we summarize how the antioxidant type and concentration influence the oxidative aging behavior. As the $[AH]_0$ increases, both $t_{\text{ind}}$ and $t_{\text{react}}$ become longer. Moreover, the smaller BHT-type antioxidant provides a more pronounced stabilization effect than the larger Irganox1010-type antioxidant. We found that this behavior arises from the local coordination structure around the BHT-type antioxidant and its higher diffusivity stemming from its low molecular weight. Because volatilization of antioxidants is not included in the present simulations, the results primarily reflect the intrinsic nature of the antioxidants in isolated systems or in bulk specimens where migration can be neglected. Although the current framework cannot account for volatilization, it should still be useful for predicting behavior in bulk-like systems and in situations where volatilization is limited, such as photo-oxidation at or below room temperature or oxidation in glassy polymers with small free volume. These insights into the origins of the stabilization effects—namely, the roles of the local static structure around antioxidant molecules and their dynamics—could only be obtained by explicitly simulating polymer dynamics as done in this study, because the dynamics of macroradicals such as POO· are governed by the motion of the polymer segments.

As a key direction for future development, the simulation could be upgraded to incorporate mass transport by coupling multiple simulation cells within a grand-canonical framework. The present simulation setup is well suited to serve as an important mesoscale simulation module for such an extension. In this study, we considered only primary antioxidants. However, it is technically feasible to incorporate secondary antioxidants into the present framework. Extending the method to enable simulations of the well-known synergistic effects between primary and secondary antioxidants [12] would be an important next step

## 4. CONCLUSION

This study examined how the antioxidant type and the initial antioxidant-group concentration control oxidative aging behavior in polymer melts. Increasing the initial antioxidant-group concentration extended both the induction time, $t_{\text{ind}}$, and the subsequent reaction time scale, $t_{\text{react}}$, indicating a systematic retardation of oxidative aging with increasing antioxidant availability. A clear difference was also observed between the two antioxidant topologies considered. The BHT-type antioxidant yielded stronger stabilization than the Irganox1010-type antioxidant under comparable antioxidant-group concentrations. This contrast cannot be attributed solely to concentration effects. Instead, it is linked to the local environment of the antioxidant groups and their mobility in the melt. In particular, the BHT-type antioxidant exhibits a more favorable local coordination with polymer segments and higher diffusivity associated with its smaller molecular size, which increases the likelihood of intercepting reactive macroradicals such as POO·. In addition, this paper examines how antioxidant concentration and molecular architecture affect the spatial heterogeneity of oxidative aging. We found that a higher antioxidant concentration mitigates the development of heterogeneous degradation. Reflecting the difference in POO· scavenging efficiency, the BHT-type antioxidant suppresses heterogeneity more effectively than the Irganox1010-type antioxidant. These results demonstrate that polymer-dynamics simulations can provide insight into the intrinsic nature of additive stabilization and the mechanistic origins of its effectiveness.

**Author Information**

**Corresponding Authors**

Takato Ishida - Department of Materials Physics, Nagoya University, Furo-cho, Chikusa, Nagoya 464-8601, Japan; orcid. Org/0000-0003-3919-2348; E-mail: ishida@mp.pse.nagoya-u.ac.jp

**Authors**

Emmanuel Richaud - Laboratoire PIMM, Arts et Metiers Institute of Technology, CNRS Cnam, HESAM Universite, 151 boulevard de l'Hopital, Paris 75013, France; orcid. Org/ 0000-0002-5315-2599; E-mail: emmanuel.richaud@ensam.eu

**Conflict of interest**

The author declares no competing interests.

**Acknowledgment**

This work was supported by JSPS KAKENHI Grant Numbers 24K20949, "Nagoya University High Performance Computing Research Project for Joint Computational Science" in Japan, CCI holdings Co., Ltd., Toukai Foundation for Technology, CASIO SCIENCE PROMOTION FOUNDATION and The Hibi Science Foundation.

induction period, (n.d.).